\documentclass[12pt]{article}
\usepackage{amsmath}
\usepackage{longtable}
\usepackage{graphicx,psfrag,epsf}
\usepackage{enumerate}
\usepackage{natbib}
\usepackage{setspace}
\usepackage{hyperref} 
\usepackage{authblk}
\usepackage{float}
\usepackage{siunitx}
\usepackage{hyperref}
\usepackage{booktabs}
\usepackage{subcaption}
\DeclareUnicodeCharacter{2212}{-}
\usepackage{url} 
\usepackage{bm}
\usepackage{ifthen} 
\usepackage{rotating}   
\usepackage{booktabs}
\usepackage{array}      
\usepackage[T1]{fontenc}
\usepackage[utf8]{inputenc} 
\usepackage{microtype}
\newcommand{\pkg}[1]{\textup{#1}}  

\usepackage{siunitx}

\usepackage{amssymb}
\newcommand{\airbnbp}{Airbnb}

\makeatletter
\newcommand*{\addFileDependency}[1]{
  \typeout{(#1)}
  \@addtofilelist{#1}
  \IfFileExists{#1}{}{\typeout{No file #1.}}
}
\makeatother

\def\spacingset#1{\renewcommand{\baselinestretch}%
{#1}\small\normalsize} \spacingset{1}

\begin{document}

\title{\bf Slomads Rising: Stay-Length Shifts in Digital-Nomad Travel, United States 2019–2024}
\author[1]{Harrison Katz}
\author[2]{Erica Savage}

\affil[1]{Data Science, Forecasting, \airbnbp}
\affil[2]{Finance, Forecasting, \airbnbp}

\maketitle

\begin{abstract}
\textbf{Background.} Length of stay, operationalized here as nights per booking (NPB), is a first-order driver of yield, labor planning, and environmental pressure. The COVID-19 pandemic and the rise of long-stay remote workers (often labeled ``slomads'', a slow-travel subset of digital nomads) plausibly altered stay-length distributions, yet national, booking-weighted evidence for the United States remains scarce. 
\textbf{Purpose.} This study quantifies COVID-19 pandemic-era and post-pandemic shifts in U.S.\ Airbnb stay lengths and identifies whether higher averages reflect (i) more long stays or (ii) longer long stays.
\textbf{Methods.} Using every U.S.\ Airbnb reservation created between 1~January~2019 and 31~December~2024 (collapsed to booking-count weights), the analysis combines: weighted descriptive statistics; parametric density fitting (Gamma, log-normal, Poisson--lognormal); weighted negative-binomial regression with month effects; a two-part (logit~+~NB) model for $\ge 28$-night stays; and a monthly SARIMA\((0,1,1)(0,1,1)_{12}\) with COVID-19 pandemic-phase indicators.
\textbf{Results.} Mean NPB rose from 3.68 pre‑COVID‑19 to 4.36 during restrictions and then stabilized near 4.07 post‑2021 ($\approx$ 10\% above 2019); the booking‑weighted median shifted permanently from 2 to 3 nights. A two‑parameter log‑normal fits best by wide AIC/BIC margins, consistent with a heavy‑tailed distribution. Negative‑binomial estimates imply post‑vaccine bookings are 6.5\% shorter than restriction-era bookings, while pre‑pandemic bookings are 16\% shorter. In a two‑part (threshold) model at 28 nights, the \emph{booking share} of month‑plus stays rose from 1.43\% (pre) to 2.72\% (restriction) and settled at 2.04\% (post), whereas the \emph{conditional} mean among long stays was in the mid‑to‑high 50s ($\approx$ 55–60 nights) and varied modestly across phases. Hence higher average NPB is driven primarily by a greater \emph{prevalence} of month‑plus bookings. A seasonal ARIMA model with pandemic‑phase dummies improves fit over a dummy‑free specification (likelihood‑ratio \(=8.39\), \(df=2\), \(p=0.015\)), indicating a structural level shift rather than higher‑order dynamics.
\textbf{Contributions.} The paper provides national-scale, booking-weighted evidence that U.S.\ short-term-rental stays became durably longer and more heavy-tailed after 2020, filling a gap in the tourism and revenue-management literature. \textbf{Implications.} Heavy-tailed pricing and inventory policies, and explicit regime indicators in forecasting, are recommended for practitioners; destination policy should reflect the larger month-plus segment.
\end{abstract}


\bigskip
\noindent
Airbnb; Nights per booking; COVID-19; weighted statistics; negative-binomial regression; United States

\section{Introduction}\label{sec:intro}

Length of stay (LOS)---here measured as nights per booking (NPB)---shapes occupancy curves, staffing, revenue yield, and per-trip environmental footprints in the visitor economy. Even a one-night shift in mean NPB can alter revenue and labor requirements at scale.

\paragraph{Context and definitions.} 
The COVID-19 pandemic introduced two countervailing pressures: restrictions that initially suppressed travel, and large-scale remote work that enabled longer,``work‑from‑anywhere'' stays. Within the broader ``digital nomad'' segment, we use ``slomads'' to denote slow-travel remote workers who favor longer, deeper month-plus residencies. Industry reports and case studies have highlighted a growing month-plus share on peer-to-peer platforms; however, national, booking-weighted evidence for the United States is limited.

\paragraph{Research gap.} 
Prior research documents LOS determinants and heavy-tailed stay patterns in peer-to-peer rentals, but most analyses are destination-specific or pre-pandemic. There is little national-scale evidence quantifying whether the observed COVID-19 pandemic-era elongation reflects (i) more long stays or (ii) longer long stays among those already staying long.

\paragraph{Objective, research questions, and hypotheses.}
\textbf{Objective:} quantify structural changes in U.S.\ Airbnb stay lengths during and after the pandemic and decompose the mechanism underlying higher averages.
We address four questions:
\begin{enumerate}
  \item[\textbf{RQ1}] Did the mean/median NPB shift persistently after 2020?
  \item[\textbf{RQ2}] Did the stay-length distribution become heavier-tailed?
  \item[\textbf{RQ3}] Is the higher average driven by a higher \emph{prevalence} of month-plus bookings or by \emph{longer} long stays?
  \item[\textbf{RQ4}] After controlling for seasonality and short-run dynamics, do pandemic-phase indicators explain the post-2020 level shift?
\end{enumerate}
Corresponding hypotheses:
\textbf{H1} (higher central tendency post-2020), \textbf{H2} (log-normal fits better than light-tailed rivals),
\textbf{H3} (prevalence increase dominates conditional duration), and
\textbf{H4} (phase indicators, not higher-order ARMA terms, account for the level shift).

\paragraph{Approach and contributions.}
This study analyzes all U.S.\ Airbnb reservations created from 2019 through 2024 using booking-count weights to recover population frequencies. It combines weighted density estimation, negative-binomial and two-part models, and SARIMA with phase indicators to isolate structural change and its mechanism. The contribution is twofold: (i) national, booking-weighted quantification of post-2020 LOS shifts; (ii) a clear decomposition showing that higher averages arise primarily from a larger month-plus segment.

\paragraph{Roadmap.}
Section~\ref{sec:lit} reviews related work. Section~\ref{sec:methods} details data and methods. Section~\ref{sec:results} presents results. Sections~\ref{sec:discussion}--\ref{sec:conclusion} interpret implications, limitations, and future research, and conclude.

\section{Literature Review}\label{sec:lit}

\noindent\textbf{Abbreviations.} LOS = length of stay; NPB = nights per booking; P2P = peer-to-peer accommodation.

This review synthesizes five strands of scholarship relevant to stay-length dynamics on short-term-rental platforms: (i) classic determinants and long‑run LOS trends; (ii) segmental heterogeneity and the heavy tail of longer trips; (iii) peer‑to‑peer accommodation, platform features, and amenity salience; (iv) COVID-19 pandemic-era remote work and the rise of long‑stay remote workers within the digital‑nomad space; and (v) statistical modeling frameworks for heavy‑tailed count outcomes and regime shifts. Table~\ref{tab:litreview} summarizes representative studies, methods, and gaps that motivate the present analysis.

\subsection{Classic determinants and long‑run LOS trends}
Early and pre‑COVID-19 pandemic empirical work documents that LOS in mature destinations has, in many cases, gradually shortened under the combined influence of falling transport costs, modular itineraries, and a ``short‑break'' culture. Micro‑econometric studies link shorter stays to budget and time constraints, with heterogeneity by purpose of trip and market segment \citep{AlegrePou2006,GokovaliBaharKozak2007,MartinezGarciaRaya2008,deMenezesMonizVieira2008,SalmasiCelidoniProcidano2012,Thrane2012,YangZhang2015,Chiou01012020}. Cross‑market syntheses report that younger and budget‑constrained travelers tend to split a fixed holiday budget across multiple brief trips rather than one extended vacation \citep{GosslingScottHall2018}. These literatures establish both a long‑run tendency toward shorter average stays in some contexts and a strong role for economic constraints and trip purpose in shaping LOS.

\subsection{Segmental heterogeneity and the heavy tail of longer trips}
A parallel literature emphasizes segments that routinely book longer stays—senior travelers, long‑haul visitors, family parties, and nature‑based tourists—producing a persistent heavy tail in the LOS distribution even where central tendencies decline \citep{AlenNicolauLosada2014,BotoGarciaBanosPino2019,BavikCorreiaKozak2021,SantosGEDO2015,GomezDenizPerezRodriguez2019}. This strand underscores that distributional shape matters for management: a small share of extended stays can account for a disproportionate share of occupied nights and revenue, and models assuming light tails will systematically understate utilization in the right tail.

\subsection{Peer‑to‑peer accommodation, platform features, and amenity salience}
Peer‑to‑peer accommodation differs systematically from hotels in traveler motivations and stay patterns, with surveys and behavioral evidence suggesting relatively longer stays on P2P platforms \citep{TussyadiahPesonen2016}. Platform design and listing attributes are implicated in these differences: trust cues and reputation systems \citep{ErtFleischerMagen2016}, pricing and demand elasticities \citep{WangNicolau2017,TangKimWang2019}, and the growing salience of ``livability'' amenities (kitchens, workspaces) that facilitate longer residencies \citep{ChenXie2017,PoonHuang2017}. These platform‑level features plausibly shift the composition of trip purposes toward medium‑term, everyday living away from home, thereby thickening the long‑stay tail.

\subsection{COVID-19 Pandemic‑era remote work, digital nomads, and long‑stay remote workers}
The COVID‑19 shock layered public‑health restrictions on travel while simultaneously expanding remote‑work arrangements. A growing literature links these forces to longer stays, work‑from‑anywhere itineraries, and an expanded digital‑nomad space \citep{BanosPinoBotoGarcia2021,JeonYang2021,SantosMoreira2021,ZhengLuoRitchie2021}. Foundational work defines digital nomadism as location‑independent, technology‑enabled mobility that blends work and travel \citep{Reichenberger2018,Hannonen2020_ITAT}, and the \emph{Information Technology \& Tourism} 2020 special issue places this phenomenon at the nexus of remote working, mobility, and infrastructures \citep{HermannParis2020}. Within this space, a slow‑travel subset—often labeled \emph{slomads}—prioritizes deeper month‑plus residencies, supported by ``livability'' amenities and coworkation practices \citep{Chevtaeva2021_JDMM}. On the policy side, many destinations introduced dedicated digital‑nomad visas during and after the pandemic; comparative analyses document their typologies and objectives \citep{Bednorz2024_ATR,KoskelaBeckers2024_JCPA,doi:10.1177/01979183241306367}. Recent mobilities scholarship further shows how nomad practices both align with and contest state mobility regimes, underscoring the structural (not merely episodic) nature of these shifts \citep{Mancinelli03032024,Hannonen19052025}. Complementary platform‑level evidence in adjacent behaviors includes broad contractions in booking lead times across major U.S.\ cities during 2020 \citep{KATZ2025100185} and spikes in currency‑share volatility \citep{katz2025bayesiandirichletautoregressiveconditional}. What remains limited is national, booking‑weighted evidence quantifying how much of the post‑2020 elongation in the United States reflects \emph{more} long stays versus \emph{longer} long stays among those already in the tail.

\subsection{Modeling frameworks for heavy‑tailed stay lengths and regime shifts}
Methodologically, LOS is a non‑negative, over‑dispersed outcome with salient right tails. Classic count frameworks motivate negative‑binomial over Poisson when $\mathrm{Var}(y)>\mathrm{E}(y)$, and two‑part (threshold) models cleanly separate event \emph{prevalence} from \emph{intensity} conditional on crossing a threshold \citep{CameronTrivedi1986,CameronTrivedi1998,Lambert1992,GurmuTrivedi1996,FamoyeSingh2006}. To capture heavy tails and the discrete nature of nights, tourism analytics has increasingly turned to the Poisson–lognormal and related variants, alongside continuous heavy‑tailed comparators such as the log‑normal \citep{GomezDenizPerezRodriguezReyes2020,ZhangTianNg2016}. Beyond single‑equation outcomes, recent work also models \emph{entire distributions} as compositional time series; in particular, Bayesian Dirichlet autoregressive moving‑average models can forecast the full vector of bin shares (e.g., lead‑time histograms), a strategy that is directly applicable when managers need dynamics for binned stay‑length shares (such as the month‑plus segment) rather than only a mean \citep{KatzBruschWeiss2024_IJF}. At the monthly aggregation, seasonal ARIMA with exogenous regressors (dynamic regression/ARIMAX) is a standard approach for identifying level shifts via intervention dummies while controlling for autocorrelation and seasonality \citep{HyndmanAthanasopoulos2021,Box01031975,HyndmanAthanasopoulos2021-Dynamic}. Across these strands, there is still relatively little work that \emph{jointly} (i) estimates the entire stay‑length distribution with booking weights, (ii) decomposes tail mechanisms via a thresholded two‑part model, and (iii) tests for regime shifts in monthly means using seasonal differencing and phase indicators.

\subsection{Synthesis and research gap}
In sum, the pre‑pandemic literature documents shorter average stays in some mature markets alongside persistent heavy tails driven by specific segments. P2P platforms, through design and amenity differences, appear to amplify the tail. Pandemic‑era research and industry evidence suggest a further expansion of long‑stay remote work and digital‑nomad mobility. Yet three gaps remain salient for the U.S. context: (G1) a lack of national, booking‑weighted estimates of post‑2020 LOS shifts; (G2) limited decomposition of whether higher averages reflect \emph{more} long stays or \emph{longer} long stays; and (G3) limited model‑based evidence on whether monthly mean NPB changes reflect structural level shifts rather than higher‑order dynamics. The empirical strategy below is designed to address G1–G3 using weighted distributional fitting, negative‑binomial and two‑part models, and seasonal ARIMA with pandemic‑phase indicators.

\section{Materials and Methods}\label{sec:methods}
\subsection{Data Structure, Period, and Weighting}\label{ssec:data}

\noindent\textbf{Data source and scope.} The analysis uses internal, de-identified Airbnb reservation records for stays in the United States, aggregated to booking-count weights (no personally identifiable information). Records cover all reservations \emph{created} between 1~January~2019 and 31~December~2024.

\noindent\textbf{Period selection (2019--2024).} The window spans (i) a stable pre-pandemic baseline, (ii) the COVID-19-restriction period, and (iii) the post-vaccine phase during which long-stay behavior stabilized. The split dates (Pre to Feb~2020; Restriction Mar~2020 - ~Jun~2021; Post from July~2021) align with the national onset of restrictions and broad vaccine availability.

\noindent\textbf{Unit and weights.} Each booking contributes an integer stay length $y_i\in\mathbb{N}$ and a weight $w_i$ equal to the count of identical reservations collapsed into row $i$. Weighted statistics and likelihoods recover population frequencies while reducing computational load.

\noindent\textbf{Upper-tail exclusion.} Bookings longer than 180 nights are excluded to focus on the short- and medium-term rental product (and to avoid conflating long-term tenancy/relocation leases). This threshold limits undue influence of extreme outliers in heavy tails.

\subsection{Weighted Descriptive Statistics}\label{ssec:descriptive}
\emph{Why:} Weighted means/quantiles summarize the realized booking population without model assumptions.

For calendar month \(t\) let \(\mathcal{I}_t\) denote the index set of
bookings.  The weight‐adjusted mean, variance and empirical quartiles
are
\[
\bar{y}_t=\frac{\sum_{i\in\mathcal{I}_t}w_i y_i}
                 {\sum_{i\in\mathcal{I}_t}w_i},\qquad
s_t^2      =\frac{\sum_{i\in\mathcal{I}_t}w_i (y_i-\bar{y}_t)^2}
                 {\sum_{i\in\mathcal{I}_t}w_i},
\]
with medians and quantiles obtained by inverting the weighted empirical
CDF.

\paragraph{Illustrative example (toy numbers).}
Suppose that in month \(t\) there are three unique stay lengths once identical reservations are collapsed to booking-count weights: \(y=\{2,7,30\}\) nights with weights \(w=\{100,40,5\}\) bookings, respectively (Table~\ref{tab:toy_weights}). The weighted mean is
\[
\bar{y}_t=\frac{2\times100 + 7\times40 + 30\times5}{100+40+5}
         =\frac{630}{145}=4.34\ \text{nights},
\]
and the weighted standard deviation is
\[
s_t=\sqrt{\frac{100(2-4.34)^2 + 40(7-4.34)^2 + 5(30-4.34)^2}{145}}
   = 5.33\ \text{nights}.
\]
The weighted empirical CDF accumulates \(100/145=0.690\) of the mass at \(y=2\), \(140/145=0.966\) by \(y=7\), and \(1.000\) by \(y=30\). Hence the median and quartiles are obtained by inversion:
\[
\tilde{y}_t = 2,\qquad Q_{0.25}=2,\qquad Q_{0.75}=7.
\]
(Values shown are rounded to two decimals; this example is purely illustrative—the analysis uses the full booking population.)

\begin{table}[H]
\centering
\caption{Toy example of booking-count weighting in month \(t\).}
\label{tab:toy_weights}
\begin{tabular}{S[table-format=2.0] S[table-format=3.0] S[table-format=3.0]}
\toprule
{Stay length \(y_i\) (nights)} & {Weight \(w_i\) (bookings)} & {Contribution \(w_i y_i\)} \\
\midrule
2  & 100 & 200 \\
7  &  40 & 280 \\
30 &   5 & 150 \\
\midrule
\textbf{Total} & 145 & 630 \\
\bottomrule
\end{tabular}
\end{table}
\subsection{Parametric Density Estimation}\label{ssec:density}
\emph{Why:} To assess tail behavior and model adequacy for pricing/forecasting, we compare a light-tailed (Gamma) and two heavy-tailed families (log-normal, discrete Poisson--lognormal).
To characterize the full distribution of nights‑per‑booking (NPB) we fit
three candidate parametric families: the continuous Gamma and
log‑normal, which are standard in hospitality analytics, and the
\emph{discrete} Poisson–lognormal (PLN), whose support matches the
integer nature of NPB.  For each model we maximize the
booking‑weighted log‑likelihood
\[
\widehat{\boldsymbol\theta}
  =\arg\max_{\boldsymbol\theta}\,
     \sum_{i} w_{i}\,\log f(y_{i}\mid \boldsymbol\theta),
\]
where \(w_{i}\) is the number of identical reservations collapsed into
row \(i\).  Gamma and log‑normal fits use
\texttt{fitdistrplus::fitdist()} with \emph{frequency} weights; because
\texttt{fitdistrplus} expects integer frequencies, we rescale and round
booking weights for this step only. The PLN uses a custom weighted
likelihood optimized by Nelder–Mead (finite‑difference gradients are
unstable near the optimum for PLN).  Models are ranked by AIC and BIC;
empirical adequacy is judged by overlays of empirical and fitted
cumulative‑distribution functions.

\subsection{Negative-Binomial Regression}\label{ssec:nbreg}
\emph{Why:} Over-dispersion in stay lengths ($\mathrm{Var}> \mathrm{E}$) motivates NB over Poisson; month dummies separate seasonality from pandemic-phase effects.

Over‑dispersion (\(\operatorname{Var}(y_i)>\operatorname{E}(y_i)\))
motivates a weighted negative‑binomial GLM:
\[
\log \operatorname{E}[y_i]
  =\beta_0
   +\beta_1\mathbb{1}_{\mathrm{Post},i}
   +\beta_2\mathbb{1}_{\mathrm{Pre},i}
   +\sum_{m=2}^{12}\gamma_m\,\mathbb{1}\{\mathrm{month}(i)=m\},
\]
where January is the reference month.
Weighted likelihoods are obtained with \texttt{MASS::glm.nb()} (robust options available
in \texttt{countreg} if needed).
Incidence‑rate ratios (IRR) and Wald 95 \% confidence intervals are
reported.

\subsection{Two-Part Long-Stay Model}\label{ssec:twopart}
\emph{Why:} A two‑part (threshold) specification cleanly decomposes the mechanism: prevalence of $\ge 28$‑night bookings (logit) versus conditional duration among long stays (NB).


Long stays are defined internally at Airbnb as \(y_i\ge28\). Part 1 predicts the probability of a long stay via a weighted
logistic model
\[
\Pr(y_i\ge28)=\operatorname{logit}^{-1}\!\bigl(
  \alpha_0+\alpha_1\mathbb{1}_{\mathrm{Post},i}
          +\alpha_2\mathbb{1}_{\mathrm{Pre},i}
          +\textstyle\sum_{m}\delta_m\mathbb{1}\{\mathrm{month}=m\}\bigr).
\]
Part 2 conditions on \(y_i\ge28\) and regresses NPB on the same
covariates using a weighted negative‑binomial GLM:  
\[
\log \operatorname{E}[y_i\mid y_i\ge28]
  =\zeta_0+\zeta_1\mathbb{1}_{\mathrm{Post},i}
          +\zeta_2\mathbb{1}_{\mathrm{Pre},i}
          +\textstyle\sum_{m}\eta_m\mathbb{1}\{\mathrm{month}=m\}.
\]

\subsection{Monthly Time-Series Model}\label{ssec:ts}
\emph{Why:} A differenced SARIMA with phase indicators tests whether the shift is a structural level change, not higher-order dynamics.

Phase indicators are first‑differenced so that regressors and
response share the same order of integration.
The differenced specification is
\[
(1-\nabla)(1-\nabla_{12})\bar{y}_t
  =\beta_0+\beta_1\nabla D_{\mathrm{Post},t}
           +\beta_2\nabla D_{\mathrm{Pre},t}
           +(1-\theta B)(1-\Theta B^{12})\varepsilon_t,
\]
where \(D_{\cdot,t}\) are the undifferenced dummies.
Model comparison between the nested ARIMA specifications (with vs.\ without phase indicators) was based on the Akaike Information Criterion (AIC) and a likelihood‑ratio (LR) test; the lower‑AIC model is preferred.

\subsection{Diagnostics and Robustness}
Computations were performed in R~4.3 \citep{RCoreTeam2023} using the
\pkg{fitdistrplus} \citep{DelignetteMuller2015}, \pkg{MASS} \citep{VenablesRipley2002},
\pkg{brglm2}, \pkg{emmeans}, \pkg{matrixStats} \citep{Bengtsson2023}, \pkg{fable} and
\pkg{fabletools} \citep{OHyde2021}, and the \pkg{tidyverse} \citep{Wickham2019} suite for
data manipulation and visualization.

\section{Results}\label{sec:results}

\subsection{Pandemic–Era Changes in Nights per booking}\label{ssec:trend}
\noindent\textit{Plain-language summary.} Average stays got longer and more variable after 2020. The mean rose by about ten percent versus 2019 and the median shifted from 2 to 3 nights; dispersion also widened, indicating a thicker tail of longer bookings.

\begin{figure}[H]
\centering
\includegraphics[width=0.9\textwidth]{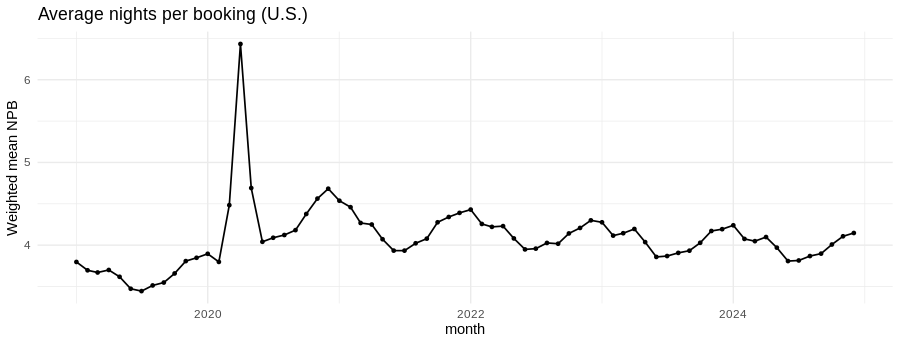}
\caption{\textbf{Mean nights per booking (NPB), U.S., by booking month, Jan~2019--Dec~2024.} 
NPB denotes nights per booking (units = nights). Values are booking‑weighted across all U.S. reservations created in each month. The mean is ~3.7 nights before the pandemic, spikes above 6 during spring 2020 lockdowns, and then stabilizes near 4.1 nights from mid‑2021 onward (about 10\% above the 2019 baseline).}
\label{fig:trend}
\end{figure}

\begin{figure}[H]
\centering
\includegraphics[width=0.9\textwidth]{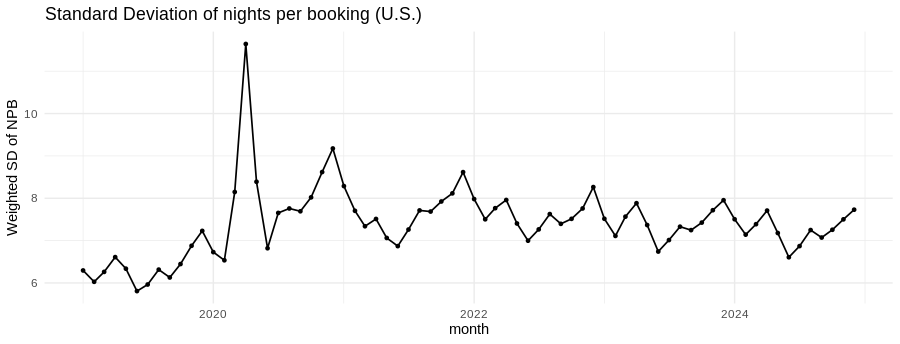}
\caption{\textbf{Standard deviation of nights per booking (NPB), U.S., by booking month, Jan~2019--Dec~2024.}
Units = nights; booking‑weighted across all U.S. reservations created in each month. Dispersion roughly doubles in spring 2020 and then settles around 7 nights from mid‑2021, with modest seasonality, indicating a persistently wider stay‑length distribution.}
\label{fig:sd}
\end{figure}

Figure~\ref{fig:trend} shows the booking-weighted mean Nights per booking (NPB) for all U.S.\ Airbnb reservations between January 2019 and December 2024.  
Mean NPB hovers around \SI{3.7}{nights} before the pandemic, surges beyond \SI{6.5}{nights} during the initial lockdown, and then settles on a post‑vaccine plateau near \SI{4.1}{nights} (phase mean \SI{4.07}{nights}).
The weighted standard deviation, presented in Figure~\ref{fig:sd}, follows the same trajectory: it doubles during spring 2020 and later stabilizes around \SI{7}{nights}, indicating a persistent widening of the stay-length distribution.

\noindent\textbf{Answer to RQ1 (H1).} Yes. The booking-weighted median shifts from 2 to 3 nights and the mean stabilizes around \SI{4.1}{nights} after 2021 (phase mean \SI{4.07}{nights}; Fig.~\ref{fig:trend}, Table~\ref{tab:desc}), confirming a persistent post‑2020 increase in central tendency.

\subsection{Phase‑wise Descriptive Statistics}\label{ssec:desc}
Weighted descriptives by pandemic phase appear in Table~\ref{tab:desc}.
The median stay length climbs from two nights pre‑COVID to three nights
during restrictions and remains at that level through 2024, confirming a
structural (not transient) elongation.

\begin{table}[H]
\caption{Booking-weighted descriptives by pandemic phase.}
\label{tab:desc}
\centering
\begin{tabular}{lcccc}
\toprule
Phase & Mean & Median & P25 & P75 \\
\midrule
Pre‑COVID           & 3.68 & 2 & 2 & 4 \\
COVID‑restrictions  & 4.36 & 3 & 2 & 4 \\
Post‑vaccine        & 4.07 & 3 & 2 & 4 \\
\bottomrule
\end{tabular}
\end{table}
\subsection{Distributional Fit}\label{ssec:distfit}
\begin{figure}[H]
\centering
\includegraphics[width=\textwidth]{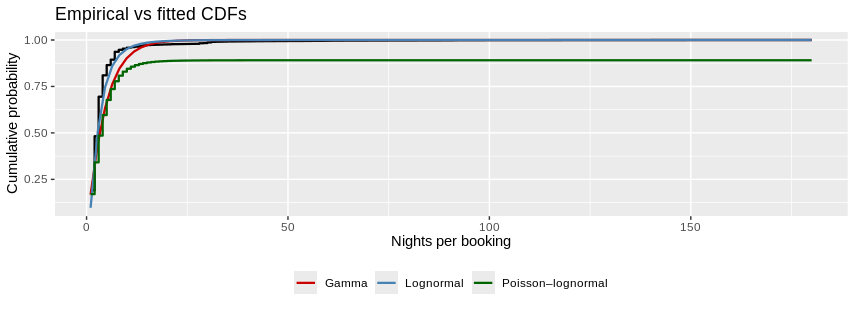}
\caption{Booking‑weighted \textbf{empirical} CDF (Cumulative Distribution Function) of nights per booking (\textbf{black step}) overlaid with fitted CDFs: \textbf{Gamma} (\textcolor{red}{red}), \textbf{log‑normal} (\textcolor{blue}{blue}), and \textbf{Poisson–lognormal} (\textcolor{green}{green}).}
\label{fig:cdf}
\end{figure}

Table~\ref{tab:gof} reports information criteria (IC) for the three candidate families. Because information criteria scale with sample size, absolute magnitudes are not directly interpretable; rank differences ((\(\Delta\)AIC/(\(\Delta\)BIC) drive model comparison.
The log‑normal achieves the best fit (AIC \(=7.71\times10^8\)); the Poisson–lognormal (PLN) is second 
(\(\Delta\)AIC \(=6.5\times10^7\)), and the Gamma is third (\(\Delta\)AIC \(=8.8\times10^7\)). 
Figure~\ref{fig:cdf} overlays the \emph{booking‑weighted empirical CDF} of nights per booking with each model’s fitted CDF. 
The log‑normal closely tracks the empirical curve across the body and upper tail; the PLN respects the integer support; 
and the Gamma under‑represents mass around the 14‑, 21‑, and 28‑night shoulders. 

\noindent\textbf{Answer to RQ2 (H2).} Yes. A two‑parameter log‑normal outperforms both the light‑tailed Gamma and the discrete PLN by wide IC margins (Table~\ref{tab:gof}); the empirical‑vs‑fitted CDF overlay is consistent with this ranking (Fig.~\ref{fig:cdf}). 
Together with the widening of the weighted standard deviation (Fig.~\ref{fig:sd}), this indicates a heavier right tail post‑2020.

\begin{table}[H]
\centering
\caption{Information criteria for booking‑weighted density fits.}
\label{tab:gof}
\begin{tabular}{lccc}
\toprule
Model & $k$ & AIC $(\times10^{8})$ & BIC $(\times10^{8})$\\
\midrule
Log‑normal        & 2 & 7.71 & 7.71\\
Poisson–lognormal & 2 & 8.36 & 8.36\\
Gamma             & 2 & 8.59 & 8.59\\
\bottomrule
\end{tabular}
\end{table}

Having established a heavier right tail, we next ask whether the higher average arises from \emph{more} long stays or \emph{longer} long stays among those already in the tail.

\subsection{Negative-Binomial Regression}\label{ssec:nb_res}
\noindent\textit{Plain-language summary.} After controlling for month fixed effects, post‑vaccine bookings are modestly shorter than during restrictions, and both are longer than pre‑COVID. Month effects trace a clear seasonal pattern: stays are shortest in early summer and lengthen again into late fall and winter, producing a U‑shaped curve across the year.

Pandemic‑phase IRRs mirror the descriptives (Table~\ref{tab:nbirr}): post‑vaccine bookings are
6.5\% shorter than restriction‑era bookings (IRR 0.935), while pre‑COVID stays are 16.2\% shorter
(IRR 0.838). By month (January = baseline), all months are modestly shorter than January; the
largest negative deviations occur in June (–10.5\%) and July (–9.9\%), with smaller differences
in late autumn (e.g., November –1.9\%, December –0.6\%). This pattern is U‑shaped across the calendar: stay lengths decline from spring into a summer minimum, then rise again through fall and winter.

\paragraph{Seasonality interpretation.}
Shorter summer stays likely reflect a higher share of brief leisure trips clustered around school breaks and mid‑year holidays, whereas longer winter stays include extended visits and medium‑term relocations. Year‑end travel peaks and mid‑year peak travel periods amplify this pattern, even though our outcome is nights per booking rather than trip counts.

\begin{figure}[H]
\centering
\includegraphics[width=\textwidth]{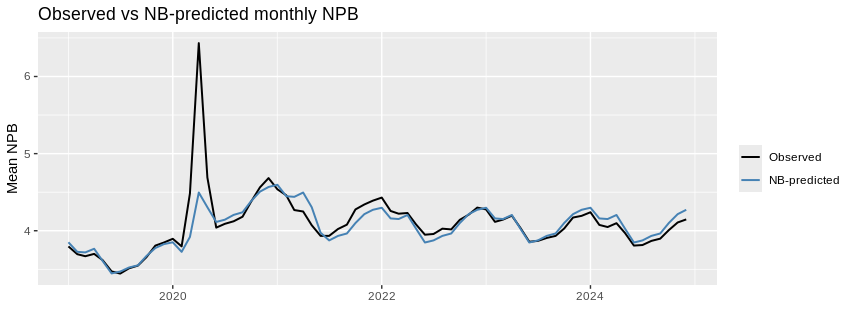}
\caption{Observed (\textbf{black}) and negative-binomial (NB) predicted (\textbf{blue}) \emph{monthly mean} nights per booking (NPB) in the United States, January 2019 to December 2024. Predictions come from a booking-weighted NB GLM with pandemic-phase indicators and month fixed effects. The model reproduces the post-2020 level shift and the U-shaped seasonal pattern; the largest residual appears during the initial lockdown spike in early 2020. Units are nights; both series are booking-weighted monthly means.}
\label{fig:pred}
\end{figure}

\begin{table}[H]
\centering
\caption{Selected incidence‑rate ratios from the NB model (January, COVID‑restriction baseline).}
\label{tab:nbirr}
\begin{tabular}{lccc}
\toprule
Covariate & IRR & 95 \% CI & Meaning\\
\midrule
Post‑vaccine & 0.935 & 0.935–0.936 & –6.5 \% vs restr.\\
Pre‑COVID    & 0.838 & 0.837–0.838 & –16.2 \% vs restr.\\
Feb          & 0.968 & 0.967–0.968 & –3.2 \% vs Jan\\
Mar          & 0.966 & 0.965–0.967 & –3.4 \% vs Jan\\
Apr          & 0.978 & 0.977–0.979 & –2.2 \% vs Jan\\
May          & 0.937 & 0.936–0.937 & –6.3 \% vs Jan\\
Jun          & 0.895 & 0.895–0.896 & –10.5 \% vs Jan\\
Jul          & 0.901 & 0.901–0.902 & –9.9 \% vs Jan\\
Aug          & 0.915 & 0.914–0.916 & –8.5 \% vs Jan\\
Sep          & 0.922 & 0.922–0.923 & –7.8 \% vs Jan\\
Oct          & 0.954 & 0.953–0.954 & –4.6 \% vs Jan\\
Nov          & 0.981 & 0.980–0.981 & –1.9 \% vs Jan\\
Dec          & 0.994 & 0.993–0.994 & –0.6 \% vs Jan\\
\bottomrule
\end{tabular}
\end{table}

Because a single‑count model conflates prevalence and conditional duration, we turn to a two‑part specification that separates the probability of a month‑plus booking from its conditional length.

\subsection{Two‑Part Long‑Stay Model}\label{ssec:twopart_res}
\noindent\textit{Headline takeaway.} The higher average is driven primarily by \emph{more} month‑plus bookings; conditional durations among long stays vary modestly across phases.

\textbf{Stage 1 (logit, $\geq$28 nights).}  
The \emph{booking share} of month‑plus stays rose from 1.43\% (Pre‑COVID) to 2.72\% (COVID‑restrictions) and settled at 2.04\% post‑vaccine (Table~\ref{tab:twopart_results}).

\textbf{Stage 2 (NB, conditional on long).}  
Among long stays, the booking‑weighted conditional mean was in the mid‑to‑high 50s: restrictions \(\approx55.0\) nights, Pre‑COVID \(\approx59.5\) nights, and Post‑vaccine \(\approx59.9\) nights.

\textbf{Combined impact.}  
The long‑segment’s contribution to occupied nights per booking equals
$\Pr(\text{long}) \times \mathrm{E}[y\mid\text{long}]$.
By phase: Pre‑COVID $0.851$, COVID‑restrictions $1.496$, Post‑vaccine $1.222$ nights per booking.
Thus the restriction era added $\approx 0.65$ nights/book relative to Pre‑COVID; the Post‑vaccine market remains above Pre‑COVID by $\approx 0.37$ nights/book but below restrictions by $\approx 0.27$.

\begin{table}[H]
\caption{Two-part long-stay model (\(\geq 28\) nights): prevalence, conditional mean, and contribution to nights per booking by phase.}
\label{tab:twopart_results}
\centering
\begin{tabular}{l S[table-format=1.2] S[table-format=2.1] S[table-format=1.3]}
\toprule
\textbf{Phase} & {\% of bookings \(\Pr(y\!\ge\!28)\)} & {\(\mathrm{E}[y \mid y\!\ge\!28]\) (nights)} & {\(\Pr\times \mathrm{E}\)} \\
\midrule
Pre‑COVID          & 1.43 & 59.5 & 0.851 \\
COVID‑restrictions & 2.72 & 55.0 & 1.496 \\
Post‑vaccine       & 2.04 & 59.9 & 1.222 \\
\bottomrule
\end{tabular}

\footnotesize\vspace{0.25em}
\emph{Notes:} Shares and conditional means are booking‑weighted. The contribution equals \(\Pr(y\!\ge\!28)\times \mathrm{E}[y \mid y\!\ge\!28]\) and is expressed in nights per booking. Values are rounded to the shown precision.
\end{table}

\noindent\textbf{Answer to RQ3 (H3).} Prevalence dominates duration. The booking share of $\geq 28$‑night stays rises from 1.43\% (pre) to 2.72\% (restriction) and stabilizes at 2.04\% (post), while conditional means among long stays are relatively stable (mid‑to‑high 50s). The net post‑vaccine uplift versus pre‑COVID is $\approx 0.37$ nights per booking.

\subsection{Time‑Series Dynamics of Monthly NPB}\label{ssec:sarima}
Because the booking‑weighted mean NPB displays both short‑run momentum
and annual seasonality, we estimated a stochastic level model with
ARIMA\((0,1,1)(0,1,1)_{12}\) errors. Two specifications were considered:
a \emph{full} model that includes the pandemic‑phase indicators as
exogenous regressors and a \emph{null} model without them. Coefficient
estimates for the full model appear in Table~\ref{tab:sarima_coef}; a
compact model‑comparison summary is given in
Table~\ref{tab:sarima_ic}.

\begin{table}[H]
\caption{Coefficient estimates for the ARIMA\((0,1,1)(0,1,1)_{12}\) with phase dummies.  Standard errors in parentheses.}
\label{tab:sarima_coef}
\centering
\begin{tabular}{lcc}
\toprule
Parameter & Estimate & s.e.\\
\midrule
Non‑seasonal MA\(_1\)   & \(-0.762\) & 0.107\\
Seasonal MA\(_{12}\)    & \(-1.000\) & 0.309\\
Post‑vaccine dummy      & \(+0.062\) & 0.192\\
Pre‑COVID dummy         & \(-1.167\) & 0.205\\
\midrule
Residual variance \(\hat{\sigma}^2\) & 0.0659 & \\
\bottomrule
\end{tabular}
\end{table}

\begin{table}[H]
\caption{Model comparison: SARIMA\((0,1,1)(0,1,1)_{12}\) with vs.\ without phase dummies.}
\label{tab:sarima_ic}
\centering
\begin{tabular}{lcc}
\toprule
 & Likelihood‑ratio \(LR\) & \(\Delta\)AIC (Full - Null) \\
\midrule
Full vs.\ Null & 8.39 \ (df = 2, \(p = 0.015\)) & \(-4.4\) \\
\bottomrule
\end{tabular}
\end{table}

\begin{figure}[H]
\centering
\includegraphics[width=\textwidth]{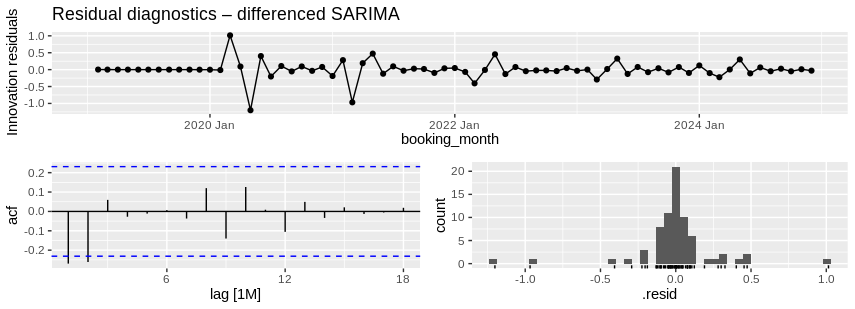}
\caption{Residual diagnostics for the differenced SARIMA model with
pandemic‑phase dummies.  Top: innovation residuals over time; 
bottom‑left: sample ACF with 95 \% bounds; 
bottom‑right: histogram of residuals.}
\label{fig:sarima_resid}
\end{figure}

Pandemic‑phase covariates improve fit: the likelihood‑ratio test rejects
the null of no phase effects (Table~\ref{tab:sarima_ic}), and AIC falls
by \(\approx4.4\) points when the dummies are included. Residual
diagnostics (Fig.~\ref{fig:sarima_resid}) show little remaining serial
structure after differencing and seasonal MA terms, supporting the view
that explicit phase indicators, rather than higher‑order ARMA terms, account
for the observed level shift in monthly NPB. The small residual
pulses are consistent with staggered state‑level tightenings and relaxations
that vary within months and are not captured by a single national phase code.

\noindent\textbf{Answer to RQ4 (H4).} Yes. Pandemic‑phase indicators improve fit over a dummy‑free SARIMA\((0,1,1)(0,1,1)_{12}\) (LR \(=8.39\), \(df=2\), \(p=0.015\)), supporting a structural level shift rather than higher‑order autoregressive dynamics.

\section{Discussion}\label{sec:discussion}
\noindent\textbf{Answers to RQ1–RQ4.}
\textit{RQ1/H1:} Post‑2020 central tendency increased and persisted: the median rose 2$\to$3 nights and the mean stabilized near 4.1 (Fig.~\ref{fig:trend}; Table~\ref{tab:desc}). At the booking level, the negative‑binomial regression confirms this pattern after month controls, with post‑vaccine stays 6.5\% shorter than the restriction era and pre‑COVID stays 16.2\% shorter (Table~\ref{tab:nbirr}).
\textit{RQ2/H2:} The stay‑length distribution is heavier‑tailed; a log‑normal fit dominates Gamma and PLN by large information‑criterion margins (Table~\ref{tab:gof}).
\textit{RQ3/H3:} Higher averages are driven primarily by a higher \emph{prevalence} of month‑plus bookings, with conditional long‑stay durations in the mid‑to‑high 50s (Section~\ref{ssec:twopart_res}).
\textit{RQ4/H4:} Monthly time‑series evidence shows that phase indicators improve fit (LR $=8.39$, $p=0.015$); higher‑order dynamics are not required.

The national‑scale evidence shows a durable structural shift in U.S. short‑term‑rental behavior: mean NPB stabilized near 4.1 (10\% above 2019) and the booking‑weighted median rose from two to three nights. Dispersion also increased, with a post‑2021 standard deviation near seven nights, confirming that the spring‑2020 spike resolved into a lasting elongation.

Distributional diagnostics reinforce this conclusion. A heavy‑tailed log‑normal density dominates Gamma and Poisson–lognormal alternatives by wide information‑criterion margins and reproduces the 14‑, 21‑, and 28‑night shoulders that the Gamma misses. Because a small but influential tail share drives utilization, light‑tailed heuristics will under‑price weekly and monthly fences.

The negative‑binomial estimates isolate phase and month effects on nights per booking while addressing over‑dispersion in the count outcome. Incidence‑rate ratios place the post‑vaccine period between the pre‑COVID baseline and the restriction era (IRR$_{\text{post}}{=}0.935$ vs restriction, IRR$_{\text{pre}}{=}0.838$), indicating that the restriction spike partially unwound but left a durable elevation. Month coefficients trace a U‑shape across the year: stays are shortest in early summer and lengthen again into late fall and winter (Table~\ref{tab:nbirr}). A plausible driver is composition shift, with more short leisure trips around school breaks and mid‑year holidays and more extended visits or temporary relocations in winter. This booking‑level pattern is consistent with the two‑part result that the post‑2020 uplift comes mainly from a higher \emph{share} of month‑plus bookings rather than longer conditional durations among those long stays.

The two‑part analysis clarifies the mechanism behind the longer average. The booking share of month‑plus itineraries increased from 1.43\% (pre‑COVID) to 2.72\% at the height of restrictions and then settled at 2.04\% post‑vaccine. Conditional durations among long stays remained in the mid‑to‑high 50s ($\approx$55–60 nights), with pre‑COVID and post‑vaccine long stays about 6–9\% longer than in the restriction phase. The combined long‑segment contribution equals $\Pr(\text{long}) \times \mathrm{E}[y\mid\text{long}]$: $0.851$ (pre), $1.496$ (restriction), and $1.222$ (post) nights per booking—an increase of $\approx 0.65$ from pre to restriction and $\approx 0.37$ from pre to post (restriction vs.\ post $\approx 0.27$).

Time‑series analysis points to the same central conclusion: explicit pandemic‑phase indicators are required to capture the level shift. In a differenced SARIMA\((0,1,1)(0,1,1)_{12}\) with phase dummies, the model outperforms a dummy‑free specification by about 4.4 AIC points, and the estimated \textit{Pre‑COVID} coefficient of \(-1.17\) nights confirms a discrete step in 2020 that persists.
The post‑2021 dummy is small and not statistically different from the restriction era once differencing and seasonality are controlled, indicating a maintained level rather than higher‑order dynamics. At the booking level, the negative‑binomial model estimates a modest 6.5\% shorter stay in the post‑vaccine period relative to the restriction era; at the monthly‑mean level, this difference is not detectable after differencing. Both approaches agree on the substantive point: post‑2021 stays remain well above pre‑pandemic levels, and phase effects dwarf month‑to‑month seasonality.

These findings have clear practical implications. Hosts and revenue managers should replace light‑tailed heuristics (for example, Gamma) with heavy‑tailed priors (for example, log‑normal) when setting weekly and monthly discounts, to avoid underpricing the now larger month‑plus segment. Platform reporting should complement booking counts with nights sold, because longer stays raise inventory‑day utilization more than they raise bookings. The persistent shift toward longer stays also supports mixed short‑term and medium‑term investment strategies. Finally, an average uplift of \(\Delta\) nights per booking implies roughly a \(\Delta/\bar{y}\) proportional increase in occupied nights (and, with the average daily rate held constant, in revenue per booking), which is why even small movements in mean NPB matter at scale.

These patterns are consistent with evidence that peer‑to‑peer accommodation use alters trip patterns relative to traditional lodging \citep{TussyadiahPesonen2016}; and with pandemic‑era studies linking remote work and digital‑nomad mobility to changes in lengths of stay and travel preferences \citep{BanosPinoBotoGarcia2021,SantosMoreira2021,ZhengLuoRitchie2021,Salon_2021}. The preference for deeper, month‑plus itineraries among slow‑travel remote workers (slomads) is consistent with amenity‑ and coworkation‑driven mechanisms \citep{Chevtaeva2021_JDMM}. Our heavy‑tailed fit aligns with work modeling tourists’ length of stay using log‑normal duration models and discrete heavy‑tailed counts (including zero‑truncated Poisson–lognormal), reinforcing the managerial point that light‑tailed assumptions can under‑represent the right tail. \citep{Chiou01012020, BotoGarciaBanosPino2019, GomezDenizPerezRodriguez2019, GomezDenizPerezRodriguezReyes2020}.

\section{Limitations and Future Research}\label{sec:limits}
This study operates on de-identified, booking-weighted records lacking host identifiers, prices, and cancellations, so it cannot quantify price elasticities, rate fences, or host-level heterogeneity. Our coarse three‑phase coding captures the national‑level discontinuity but not the staggered state‑by‑state lockdowns, partial re‑openings, and rollbacks that occurred within months. Richer exogenous controls such as mobility measures could explain additional month‑to‑month variation in a dynamic‑regression extension, but our primary objective here is to identify and quantify the structural level shift in stay lengths. Future work should (i) integrate price and cancellation data to link LOS shifts to revenue/yield; (ii) study regional and policy heterogeneity; and (iii) test natural experiments (e.g., return-to-office mandates, visa changes) to strengthen causal attribution.

\section{Conclusion}\label{sec:conclusion}
These results directly answer RQ1–RQ4 and support H1–H4: higher post‑2020 central tendency (RQ1/H1), a heavier‑tailed distribution (RQ2/H2), a prevalence‑driven mechanism (RQ3/H3), and a phase‑indicator‑captured level shift (RQ4/H4). Using national, booking-weighted Airbnb data for 2019--2024, this study shows that U.S.\ short-term-rental stays became durably longer and more heavy-tailed after 2020. The median rose from two to three nights and the mean stabilized about ten percent above the 2019 baseline. A two-part decomposition reveals that the higher average stems chiefly from a larger month-plus segment, not longer durations among existing long-stay guests, while time-series evidence points to a structural level shift. 
\textit{Managerial implications:} adopt heavy-tailed priors when setting weekly/monthly discounts; incorporate regime indicators in forecasting; and track nights sold (not only bookings) when planning inventory and staffing. 
\textit{Policy implications:} destinations and regulators should recognize the larger medium-term segment in zoning, taxation, and housing debates. 
\textit{Research implications:} integrate prices, cancellations, and geography to unpack mechanisms and heterogeneity.

\section*{Acknowledgement}
The authors thank Sean Wilson, Liz Medina, Jess Needleman, Jackson Wang, Jenny Cheng, and Peter Coles for helpful discussions, and Ellie Mertz and Adam Liss for championing the research. Special thanks are
extended to Lauren Mackevich for their indispensable operational
support.

\section*{Conflict of Interest}
The authors declare the following financial interests/ personal relationships which may be considered as potential competing interests: Harrison Katz reports a relationship with Airbnb that includes: employment. Erica Savage reports a relationship with Airbnb that includes: employment.

\bibliographystyle{apalike}

\bibliography{references}

\appendix
\begin{sidewaystable}[htbp]
\scriptsize
\centering
\caption{Selected literature on length of stay (LOS), peer-to-peer accommodation, pandemic-era travel, digital-nomadism/slomads, policy responses, and modeling approaches. Representative studies are listed; see text for fuller discussion.}
\label{tab:litreview}
\begin{tabular}{p{3.2cm} p{5.5cm} p{4.5cm} p{4.5cm} p{5.5cm}}
\toprule
\textbf{Strand} & \textbf{Representative studies} & \textbf{Setting/data} & \textbf{Methods} & \textbf{Key findings \& gap} \\
\midrule
LOS determinants \& trends 
& \citep{AlegrePou2006,GokovaliBaharKozak2007,MartinezGarciaRaya2008,deMenezesMonizVieira2008,SalmasiCelidoniProcidano2012,Thrane2012,YangZhang2015,GosslingScottHall2018} 
& Mature destinations; household/trip data 
& Micro-econometrics; GLM 
& Short-break trend; budget/time constraints shorten LOS. Mostly pre-pandemic; limited national U.S. booking-weighted evidence. \\

Segmental heavy tails 
& \citep{AlenNicolauLosada2014,BotoGarciaBanosPino2019,BavikCorreiaKozak2021,SantosGEDO2015,GomezDenizPerezRodriguez2019} 
& Seniors, long-haul, family, nature-based 
& Discrete choice; GLM 
& Specific segments drive longer stays and right tails. Need distribution-aware management. \\

P2P accommodation 
& \citep{TussyadiahPesonen2016} 
& P2P vs.\ hotels 
& Surveys; behavioral 
& P2P associated with relatively longer stays. Destination-specific; pre-/early-pandemic. \\

Platform features \& amenities 
& \citep{ChenXie2017,ErtFleischerMagen2016,WangNicolau2017,TangKimWang2019,PoonHuang2017} 
& Platform-level attributes 
& Demand/price models; experiments 
& Trust, reputation, and ``livability'' amenities correlate with longer residencies. Mechanisms suggest a thicker tail. \\

\textbf{Digital-nomad definitions \& field framing} 
& \citep{Reichenberger2018,Hannonen2020_ITAT,HermannParis2020} 
& Conceptual/definitional; integrative field overviews 
& Conceptual syntheses; editorial framing 
& DN as location-independent, tech-enabled mobility blending work \& travel; frames remote work/mobility/infrastructure nexus. \emph{Gap:} connect definitional anchors to national, booking-weighted platform evidence. \\

\textbf{Pandemic-era remote work \& long-stay remote workers (slomads)} 
& \citep{BanosPinoBotoGarcia2021,ZhengLuoRitchie2021,Chevtaeva2021_JDMM} 
& COVID-era travel panels/case studies; coworkation contexts 
& Panels; case studies; survey/qualitative 
& Remote work enables month-plus itineraries; coworkation/amenities facilitate deeper stays. \emph{Gap:} U.S. national, booking-weighted decomposition of ``more'' long stays vs ``longer'' long stays. \\

\textbf{Policy environment: digital-nomad visas} 
& \citep{Bednorz2024_ATR,KoskelaBeckers2024_JCPA,doi:10.1177/01979183241306367} 
& Cross-country visa/policy texts and programs 
& Comparative policy analysis; typology; objectives mapping 
& Rapid proliferation of DN visas with diverse rationales (tourism, talent, fiscal). \emph{Gap:} link visa regimes to platform stay-length distributions and operations at scale. \\

\textbf{Structural/critical mobilities} 
& \citep{Mancinelli03032024,Hannonen19052025} 
& Mobilities \& tourism-geography perspectives 
& Qualitative/critical; conceptual synthesis 
& Nomad practices negotiated with/against state regimes; emergent geographies suggest structural (not episodic) shifts. \emph{Gap:} connect to quantitative LOS outcomes. \\

Platform-level structural shifts 
& \citep{KATZ2025100185,katz2025bayesiandirichletautoregressiveconditional} 
& U.S.; platform aggregates 
& Normalized distances; volatility analysis 
& Booking windows contracted and currency shares became more volatile in 2020, indicating structural change beyond LOS. \\

Modeling frameworks 
& \citep{CameronTrivedi1986,CameronTrivedi1998,Lambert1992,GurmuTrivedi1996,FamoyeSingh2006,ZhangTianNg2016,KatzBruschWeiss2024_IJF,GomezDenizPerezRodriguezReyes2020,Chiou01012020} 
& LOS/over-dispersion; heavy tails 
& NB; two-part (threshold); PLN; log-normal 
& Tools to model over-dispersion and tails. Few studies combine booking weights, tail decomposition, and regime testing. \\
\bottomrule
\end{tabular}
\end{sidewaystable}

\begin{sidewaystable}[htbp]
\scriptsize
\centering
\caption{Crosswalk from research questions and hypotheses to empirical tests and conclusions.}
\label{tab:rq_crosswalk}
\begin{tabular}{p{1.8cm} p{5.2cm} p{5.5cm} p{4.0cm} p{3.5cm}}
\toprule
\textbf{RQ / H} & \textbf{Question / Hypothesis} & \textbf{Primary evidence (model/statistic)} & \textbf{Answer} & \textbf{Where} \\
\midrule
RQ1 / H1 & Persistent post‑2020 shift in central tendency & Booking‑weighted descriptives; NB IRRs; median 2\(\to\)3; mean 3.7\(\to\)4.1 & Supported & Fig.~\ref{fig:trend}; Table~\ref{tab:desc}; Table~\ref{tab:nbirr} \\
RQ2 / H2 & Heavier‑tailed distribution; log‑normal vs rivals & AIC/BIC comparison (log‑normal vs Gamma/PLN); widened sd & Supported & Table~\ref{tab:gof}; Fig.~\ref{fig:sd} \\
RQ3 / H3 & Mechanism: more long stays vs longer long stays & Two‑part model: booking share 1.43\% (pre) $\to$ 2.72\% (restr) $\to$ 2.04\% (post); conditional means $\approx$55–60 nights; combined contributions $c_{\text{pre}}{=}0.851$, $c_{\text{restr}}{=}1.496$, $c_{\text{post}}{=}1.222$ & Supported (prevalence) & Section~\ref{ssec:twopart_res} \\
RQ4 / H4 & Phase dummies explain level shift vs higher‑order dynamics & SARIMA\((0,1,1)(0,1,1)_{12}\)+dummies; LR \(=8.39\), \(df=2\), \(p=0.015\); \(\Delta\)AIC \(\approx -4.4\) & Supported & Section~\ref{ssec:sarima} \\

\bottomrule
\end{tabular}
\end{sidewaystable}

\end{document}